\documentclass[cameraready]{Interspeech}

\title{USAD 2.0: Scaling Representation Distillation \\ for Universal Audio Understanding}

\author[affiliation={1}, orcid=0000-0002-1690-2610, correspondingauthor]{Heng-Jui}{Chang}
\author[affiliation={1}]{Alexander H.}{Liu}
\author[affiliation={2}]{Saurabhchand}{Bhati}
\author[affiliation={2}]{Mrudula}{Athi}
\author[affiliation={2}]{~~~~~~~~~~~~~~~~~~~~~~~Anton}{Ratnarajah}
\author[affiliation={2}]{Amit}{Chhetri}
\author[affiliation={1}]{James}{Glass}

\address{
    $^1$MIT CSAIL, USA ~~
    $^2$Amazon, USA
}

\email{hengjui@mit.edu}


\usepackage{comment}

\usepackage{cite}
\usepackage{amsmath,amssymb,amsfonts}
\usepackage{algorithmic}
\usepackage{graphicx}
\usepackage{textcomp}
\usepackage[table]{xcolor}

\usepackage{hyperref}
\usepackage{multirow}
\usepackage{threeparttable}
\usepackage{subcaption}
\usepackage{booktabs}
\usepackage{graphicx}
\usepackage{amssymb,amsmath}
\usepackage{adjustbox}
\usepackage{enumitem}
\usepackage{bbding}
\usepackage{stfloats}
\usepackage{url}
\usepackage{cite}

\usepackage{etoolbox}
\apptocmd{\normalsize}{%
  \setlength{\abovedisplayskip}{4pt plus 2pt minus 2pt}%
  \setlength{\belowdisplayskip}{4pt plus 2pt minus 2pt}%
  \setlength{\abovedisplayshortskip}{2pt plus 2pt}%
  \setlength{\belowdisplayshortskip}{3pt plus 2pt minus 2pt}%
}{}{}

\begin{document}

\maketitle

\begin{abstract}
    Audio encoders are critical to modern audio applications as large language models~(LLMs) increasingly rely on a single encoder for diverse inputs.
    While self-supervised learning~(SSL) has yielded strong domain-specific encoders like speech or music experts, multi-domain approaches like USAD and SPEAR remain limited in coverage and evaluation.
    Recent studies also suggest supervised encoders align better with audio LLMs.
    We present USAD~2.0, a universal encoder integrating knowledge from both SSL and supervised foundation models.
    USAD~2.0 introduces domain-aware distillation to address teacher mismatch, extends coverage to the music domain, and adds second-stage supervised distillation for downstream use.
    We further scale the model to one billion parameters via depth scaling.
    Experiments show USAD~2.0 achieves strong or state-of-the-art performance across probing and LLM-based evaluations.\footnote{\url{https://hf.co/collections/MIT-SLS/usad2}}
\end{abstract}

\section{Introduction}
\label{sec:intro}

Audio encoders have been extensively explored for applications ranging from automatic speech recognition~(ASR) to audio codecs~\cite{baevski2020wav2vec2,chang2025dcspin}.
These encoders transform raw waveforms into compact representations, allowing downstream models to access information from audio signals.
A widely adopted approach is self-supervised learning~(SSL) on large unlabeled datasets, which provides fine-grained features and reduces the reliance on annotated data~\cite{yang2024large}.
However, most SSL models are curated for single-domain usage.
E.g., WavLM~\cite{chen2022wavlm} excels at speech tasks but struggles with out-of-domain audio such as environmental soundscapes.
Similar limitations can be observed in general audio~\cite{chen2022beats,chen2024eat,dinkel24bdasheng,li2024atst,alex2025sslam} and music~\cite{li2023mert,won2024musicfm,zhu2025muq} SSL models.

With recent advances in audio large language models~(LLMs), there is a growing need for strong audio frontends that produce high-quality embeddings across domains, motivating multi-domain audio SSL models.
Universal Speech and Audio Distillation~(USAD)~\cite{chang2025usad} proposes layer-wise distillation to aggregate knowledge from speech and general-audio SSL encoders.
In parallel, Wei et al. distill knowledge from speech and music experts~\cite{wei2025multi-distillation}, and SPEech and Audio Representations~(SPEAR) distills from multi-codebook vector-quantized SSL models~\cite{yang2025spear}.
Nevertheless, these models are primarily evaluated via probing tasks and do not simultaneously cover speech, general audio, and music domains.

Meanwhile, recent studies suggest that \textit{supervised} audio encoders can be particularly effective for audio LLMs, audio retrieval, and speech codecs~\cite{chu2024qwen2audio,dinkel2025midashenglm,song2025stabletoken,vyas2026peav}.
E.g., the encoder in Audio Flamingo 3~\cite{goel2025af3} is initialized from Whisper Large~\cite{radford2022whisper} and then fine-tuned with joint audio captioning and ASR objectives.
With explicit alignment to target applications, such encoders are more likely to succeed as frontends for multimodal LLMs.

\begin{figure}[t]
    \centering
    \includegraphics[width=\linewidth]{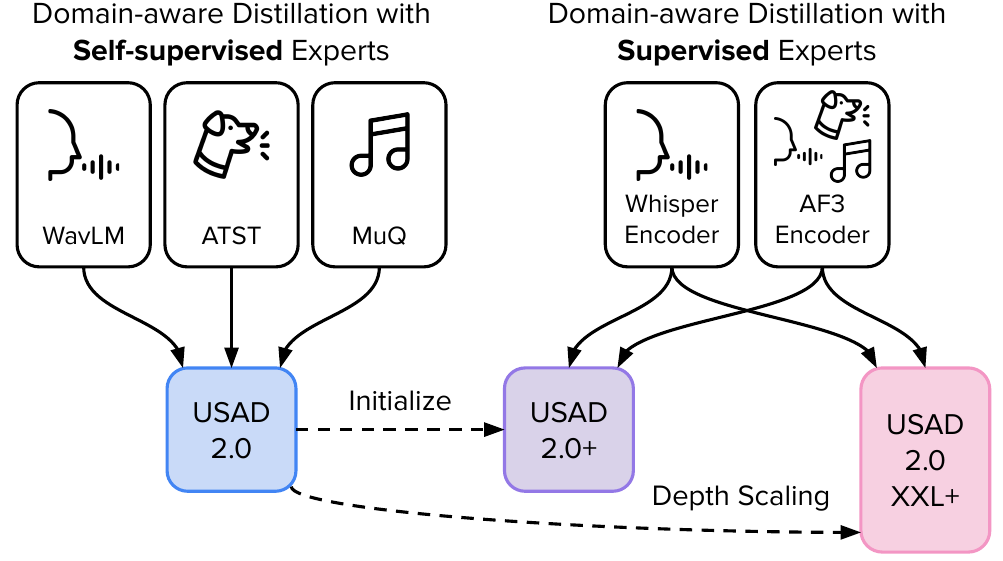}
    \vspace{-20pt}
    \caption{
        Proposed USAD 2.0.
        Domain-aware distillation from three SSL experts establishes a strong foundation.
        Next, supervised experts are distilled to the first stage-initialized encoder.
        Finally, depth scaling is applied to increase model capacity.
    }
    \label{fig:usad2-overview}
    \vspace{-10pt}
\end{figure}

In this paper, we build a universal audio encoder that extracts useful representations across multiple audio domains and tasks by distilling from both SSL and supervised audio foundation models.
We propose USAD 2.0, which builds on USAD~\cite{chang2025usad} to provide a practical, systematically evaluated framework for integrating domain-specialized audio encoders.
As shown in Fig.~\ref{fig:usad2-overview}, we first introduce domain-aware distillation, which accounts for whether a teacher matches the input domain.
We then incorporate a music teacher along with large-scale music datasets.
Next, we propose USAD 2.0+ via second-stage distillation from supervised state-of-the-art teachers to align the encoder with audio LLM applications.
Finally, we scale USAD 2.0+ to one billion parameters by reducing temporal resolution and scaling depth with minimal cost.
USAD 2.0 achieves superior performance on both probing and LLM-based evaluations across diverse audio domains, demonstrating the effectiveness of the proposed framework as a universal audio encoder.
Comprehensive ablation experiments and visualization justify the efficacy of the proposed techniques.

\section{Methods}
\label{sec:method}

\subsection{Recap: USAD}
\label{sec:method-usad}

Universal Speech and Audio Distillation~(USAD) distills knowledge from two SSL models, one specializing in speech and the other in general audio, into a single encoder for universal audio understanding~\cite{chang2025usad}.
USAD uses layer-to-layer knowledge distillation, motivated by the observation that different information types, such as speech content and environmental sounds, are encoded across the hidden layers of SSL models~\cite{chang2024colld}.
Although USAD used only two SSL teachers, we generalize the formulation to $M$ teachers.
The training loss is the average of per-teacher distillation losses $\mathcal{L}_m$, each decomposed into layer-wise terms $\mathcal{L}_{m,k}$:
\begin{equation}
    \mathcal{L}_{\text{USAD}} = \frac{1}{M}\sum_{m=1}^M \mathcal{L}_m = \frac{1}{MK}\sum_{m=1}^M \sum_{k=1}^K \mathcal{L}_{m,k} ,
\end{equation}
where $K$ is the number of layers from which the student distills for each teacher model.
Each $\mathcal{L}_{m,k}$ follows DistilHuBERT~\cite{chang2022distilhubert} by maximizing the similarity between the student and teacher hidden representations.
By directly learning the behavior of SSL experts, USAD achieves balanced performance across multiple tasks and data domains.
Building on this success, we propose USAD~2.0, which incorporates improved distillation~(Sec.~\ref{sec:method-usad2}) and scaling techniques~(Sec.~\ref{sec:method-scaling}) to support a broader range of audio understanding applications.

\subsection{USAD 2.0}
\label{sec:method-usad2}

\subsubsection{Domain-aware Distillation}
\label{sec:method-usad2-dcd}

This section introduces domain-aware distillation to improve USAD.
Each SSL teacher specializes in a specific audio domain, but USAD weights all teacher losses equally, regardless of the input.
To encourage higher-quality representations, we upweight the loss when the input domain matches the corresponding teacher domain.
Assuming the $M$ teachers each specialize in a unique domain, the loss for an instance from domain $m_{\text{data}}$ is
\vspace{-5pt}
\begin{equation}
    \mathcal{L}_{\text{USAD 2.0}} = \sum_{m=1}^M w_m(m_{\text{data}}) \mathcal{L}_m,
    \label{eq:loss-usad2}
\end{equation}
where $w_m(m_{\text{data}})$ scales the contribution of the $m$\textsuperscript{th} teacher.
We introduce a scaling factor $\alpha > 1$ to control the ratio between matched and mismatched domains.
Enforcing $\sum_{m=1}^M w_m(m_{\text{data}})=1$, we define
\begin{align}
    w_m(m_{\text{data}}) = \left\{
    \begin{array}{ll}
        \frac{\alpha}{\alpha + M - 1} & , m = m_{\text{data}} \\
        \frac{1}{\alpha + M - 1} & , m \neq m_{\text{data}}
    \end{array}
    \right.
\end{align}
When $\alpha = 1$, the weights reduce to $\frac{1}{M}$ for all teachers.
If $m_{\text{data}}$ is unknown, we also set $w_m=\frac{1}{M}$.
Unlike~\cite{wei2025multi-distillation}, which effectively takes $\alpha \rightarrow \infty$, our \textit{soft} weighting still allows distillation from mismatched teachers.
This is beneficial when domains share structure.
E.g., since speech often appears in mixed audio, distilling from a speech teacher can help the student acquire denoising capability.
Thus, mismatched teachers remain active with smaller weights, allowing the student to retain cross-domain cues while still emphasizing the most relevant expert for each input domain.

\subsubsection{Music Domain Expert}
\label{sec:method-usad2-music}

Empirically, USAD underperforms music SSL models on music-centric tasks such as genre and key classification, likely due to the lack of music-domain supervision.
Given the growing importance of music-focused SSL methods and applications~\cite{li2023mert,won2024musicfm,zhu2025muq,ghosh2025mf}, we introduce a music-domain expert and additional music audio data for USAD~2.0 to distill from.
Combined with domain-aware distillation, this gives USAD~2.0 a broader and more diverse skill set.

\subsection{Scaling USAD 2.0}
\label{sec:method-scaling}

\subsubsection{Second-stage Distillation with Supervised Experts}
\label{sec:method-scaling-supervised}

Recent progress in audio LLMs has highlighted the effectiveness of audio encoders pre-trained with supervised objectives like ASR.
In particular, many audio encoders are fine-tuned from Whisper's encoder~\cite{radford2022whisper,chu2024qwen2audio,goel2025af3,ghosh2025mf}.
Hence, we propose USAD~2.0+ via second-stage distillation from state-of-the-art supervised audio encoders.
We first identify the strongest experts using probing and LLM-based evaluations: the Whisper Large encoder for multilingual speech~\cite{radford2022whisper} and the Audio Flamingo 3 encoder for general audio understanding~\cite{goel2025af3}.
We then initialize USAD~2.0+ from the SSL-distilled student and distill from the final layers of both supervised teachers.
This stage aligns USAD~2.0 with audio LLMs while preserving the fine-grained representations characteristic of SSL pre-training.

\subsubsection{Efficient Model Size Scaling}
\label{sec:method-scaling-size}

Scaling the audio encoder can improve downstream performance by increasing model capacity, but incurs substantially higher computational cost.
We therefore propose two simple approaches to scale USAD~2.0: temporal resolution reduction and depth scaling.
First, since the sequence length processed by self-attention dominates training and inference cost, we reduce the feature framerate from 50Hz to 25Hz with a 2$\times$ CNN feature extractor stride.
Although the temporal resolution is reduced, increasing the number of layers and hidden dimensions can still improve the encoder's overall capacity and capability.
Second, we reuse the weights of a pre-trained USAD~2.0 model, apply depth scaling, and train the expanded model for only a few more updates.
Specifically, we scale our XLarge model from 32 to 48 layers with depth up-scaling~\cite{kim2024solar} by copying and stacking the first and last 24 layers.
These methods avoid training large models from scratch and enable scaling USAD~2.0 to 1B parameters within an academic budget.

\section{Experiments}
\label{sec:exp}

\begin{table}[t]
    \centering
    \caption{
        Results on HEAR~\cite{turian2022hear}, MARBLE~\cite{yuan2023marble}, and XARES-LLM.
        All reported numbers are obtained by using only the audio encoder of each model.
        E.g., the decoder of each Whisper model is discarded.
        The best results are shown in \textbf{bold}, and the second- and third-best results are \underline{underlined}.
    }
    \label{tab:main-eval}
    \vspace{-8pt}
    \begin{adjustbox}{max width=\linewidth}
    \begin{threeparttable}
    \begin{tabular}{@{}l@{~}r@{~}c@{~}c@{~}c@{~}c@{}}
        \toprule
         & & HEAR & MARBLE~ & \multicolumn{2}{@{~}c@{}}{XARES-LLM} \\
        \cmidrule{5-6}
        Encoder & Params & Avg & Avg & Track A & Track B \\
        \midrule
        \multicolumn{4}{@{}l@{~}}{\textbf{Single-encoder SOTA}} \\
        ~~Base & $\sim$90M & 80.6 & 74.0 & 0.660 & 0.418 \\
        & &  \scriptsize \shortstack{SPEAR \\ Base~\cite{yang2025spear}} & \scriptsize \shortstack{MERT \\ 95M~\cite{li2023mert}} & \scriptsize \shortstack{SPEAR \\ Base~\cite{yang2025spear}} & \scriptsize \shortstack{Whisper \\ Small~\cite{radford2022whisper}}  \\[3pt]
        
        ~~Large & $\sim$300M & 81.8 & \textbf{77.0} & 0.691 & 0.454 \\
        & & \scriptsize \shortstack{SPEAR \\ Large~\cite{yang2025spear}} & \scriptsize \shortstack{MuQ~\cite{zhu2025muq}\\~} & \scriptsize \shortstack{SPEAR \\ Large~\cite{yang2025spear}} & \scriptsize \shortstack{Whisper\\ Medium~\cite{radford2022whisper}} \\[3pt]
        
        ~~XLarge & $\sim$600M & 82.6 & 75.1 & \underline{0.782} & 0.457 \\
        & & \scriptsize \shortstack{SPEAR \\ XLarge~\cite{yang2025spear}} & \scriptsize \shortstack{SPEAR \\ XLarge~\cite{yang2025spear}} & \scriptsize \shortstack{AF3~\cite{goel2025af3}\\~} & \scriptsize \shortstack{Whisper \\ Large~\cite{radford2022whisper}} \\[2pt]

        \midrule
        \multicolumn{5}{@{}l@{~}}{\textbf{Multi-expert Encoder (USAD 2.0 Teachers)}} \\
        ~~Self-supervised & \multirow{2}{*}{734M} & \multirow{2}{*}{82.0} & \multirow{2}{*}{\underline{76.1}} & \multirow{2}{*}{0.645} & \multirow{2}{*}{0.462} \\
        ~~~{\scriptsize (WavLM + ATST + MuQ)} \hspace{-6pt} &  \\
        ~~Supervised & \multirow{2}{*}{1274M} & \multirow{2}{*}{81.8} & \multirow{2}{*}{72.4} & \multirow{2}{*}{\textbf{0.806}} & \multirow{2}{*}{\textbf{0.685}} \\
        ~~~{\scriptsize (Whisper + AF3)} &  \\
        \midrule
        \multicolumn{4}{@{}l@{~}}{\textbf{USAD 2.0 (Self-supervised Teachers)}} \\
        ~~Small & 25M & 81.0 & 72.9 & 0.604 & 0.357 \\
        ~~Base & 97M & 81.9 & 74.1 & 0.645 & 0.442 \\
        ~~Large & 336M & \underline{82.9} & \underline{75.8} & 0.667 & 0.473 \\
        ~~XLarge & 695M & 82.5 & 75.7 & 0.708 & 0.485 \\

        \midrule
        \multicolumn{4}{@{}l@{~}}{\textbf{USAD 2.0+ (Supervised Teachers)}} \\
        ~~Large+ & 336M & \underline{84.0} & 75.1& 0.769 & 0.580 \\
        ~~XLarge+ & 695M & \textbf{84.4} & 75.0 & 0.772 & \underline{0.611} \\
        ~~XXLarge+ & 1036M & \textbf{84.4} & 75.6 & \underline{0.783} & \underline{0.624} \\
        \bottomrule
    \end{tabular}
    \end{threeparttable}
    \end{adjustbox}
    \vspace{-8pt}
\end{table}

\subsection{Setup}
\label{sec:exp-setup}

\subsubsection{USAD 2.0 Training}
\label{sec:exp-setup-usad2}

Following USAD~\cite{chang2025usad}, we create a multi-domain audio dataset by combining various multilingual speech~(116K hours)~\cite{Panayotov2015libri,kahn2020libri,pratap2020mls,ardila2020commonvoice,wang2021voxpopuli,chen2024xeus,pratap2024mms,chen2021gigaspeech,valk2021voxlingua107,cieri2004fisher,roach1998babel,conneau2023fleurs,bu2017aishell,barker2015chime3,nguyen2023expresso}, general audio~(21K hours)~\cite{aytar2016soundnet,gemmeke2017audioset,wu2023laion-audio,chen2020vggsound}, and music~(13K hours)~\cite{defferrard2016fma,bogdanov2019mtg,santana2020music4all,engel2017nsynth,law2009mtt,hawthorne2018maestro} corpora.
The domain labels are assigned according to each dataset's original purpose, and the domain-aware distillation scale $\alpha$ is set to 10 for all models.
USAD~2.0 follows the same architecture as USAD~\cite{chang2025usad}, except that the XLarge and XXLarge models use a 25Hz framerate for efficiency.
For first-stage training, USAD~2.0 distills from WavLM~\cite{chen2022wavlm}, ATST-Frame~\cite{li2024atst}, and MuQ~\cite{zhu2025muq}, respectively representing speech, audio, and music experts.
Thus, this stage directly evaluates the extension from the two-teacher setting of USAD to three SSL teachers.
Supervised distillation uses the Whisper Large-v3 encoder~\cite{radford2022whisper} and Audio Flamingo 3~(AF3) AF-Whisper~\cite{goel2025af3} as targets, distilling only the last layer of each expert due to their supervised nature.
Because AF3 is multi-domain, the losses from all domains are treated equally.
The first and second stages are trained with 600K and 50K updates, respectively.

\subsubsection{Evaluation}
\label{sec:exp-setup-eval}

We include multiple protocols to evaluate the proposed models, ranging from simple probing tasks to audio LLM evaluations.
\textbf{HEAR} is a benchmark that probes frozen SSL model representations for various tasks, covering speech, sound, and music~\cite{turian2022hear}.
\textbf{MARBLE} is a music-focused probing benchmark similar to HEAR~\cite{yuan2023marble}.
Finally, we follow \textbf{XARES-LLM}~(The Interspeech 2026 Audio Encoder Capability Challenge for Large Audio Language Models) by training a multitask audio LLM using frozen representations of audio encoders~\cite{dinkel2026interspeech}.
Track A~(classification tasks) covers keyword spotting, speaker/language identification, spoof detection, intent/emotion/sound/genre/instrument classification, and sound event detection.
Track B~(understanding tasks) includes English/Mandarin ASR and audio/music captioning.
To ensure controlled comparison of audio representations, all baselines are evaluated in an encoder-only setting, including prior universal or multi-domain encoders~\cite{chang2025usad,yang2025spear}, domain-specialized SSL models~\cite{chen2022wavlm,li2024atst,zhu2025muq}, supervised audio LLM-oriented encoders~\cite{radford2022whisper,goel2025af3,ghosh2025mf}, and multi-expert teacher toplines.

\subsection{Main Results}

Tab.~\ref{tab:main-eval} reports average scores of each benchmark.
On HEAR, the unsupervised USAD 2.0 models consistently outperform prior state-of-the-art models of comparable sizes.
Although reducing the framerate to 25Hz slightly degrades the XLarge model performance, the score remains competitive with SPEAR XLarge~\cite{yang2025spear}.
Introducing the second-stage distillation with supervised teachers~(USAD 2.0+) yields further improvements, pushing beyond the prior state-of-the-art model.

For the music-centric evaluation on MARBLE, USAD 2.0 demonstrates robust multi-domain coverage.
The unsupervised Large model surpasses both the Base and XLarge baselines, while still highly competitive with specialized, music-only models like MuQ~\cite{zhu2025muq}.
The supervised variants maintain this strong performance, indicating that aligning with supervised experts preserves fine-grained music understanding.

On XARES-LLM, USAD 2.0 exhibits highly effective scaling.
The supervised USAD 2.0+ variants match or surpass the top-performing XLarge single-encoder baselines, especially on Track B.
The most significant gains are observed in Track B~(understanding).
The results indicate the usefulness of distilling from supervised experts.

To provide a performance topline, we include the ``Multi-expert Encoder'' results, which are obtained by concatenating the outputs of the teacher models.
While these ensembles achieve high scores, they require significantly more parameters than our distilled students.
Notably, USAD 2.0 models often match or exceed these multi-expert teachers while maintaining a much smaller parameter footprint.
By integrating the fine-grained acoustic details of SSL experts with the high-level semantic alignment of supervised models, USAD 2.0 establishes a new state-of-the-art for efficient, highly capable universal audio encoders across speech, sound, and music.

\subsection{Ablation Studies}

\begin{figure}[t]
    \centering
    \includegraphics[width=\linewidth]{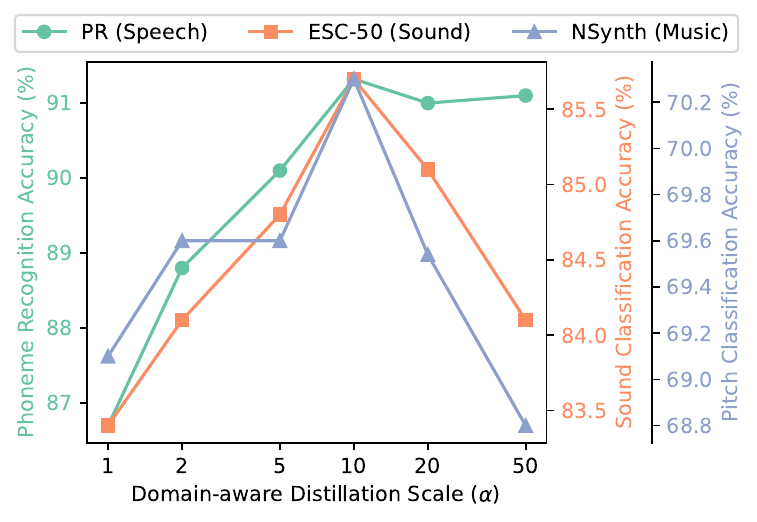}
    \vspace{-22pt}
    \caption{
        Domain-aware distillation scale vs. phoneme recognition, sound classification, and pitch classification, where $\alpha=10$ is most robust across domains.
    }
    \label{fig:domain-scale}
    \vspace{-3pt}
\end{figure}
\begin{table}[t]
    \centering
    \caption{
        Ablation studies on phoneme recognition~(PR)~\cite{yang2021superb,Panayotov2015libri}, sound classification~(ESC-50)~\cite{piczak2015esc50}, and pitch classification~(NSynth)~\cite{engel2017nsynth}.
        All models use the Small 25M-parameter backbone without fine-tuning.
    }
    \label{tab:ablation}
    \vspace{-8pt}
    \begin{adjustbox}{max width=\linewidth}
    \begin{threeparttable}
    \begin{tabular}{@{}lccc@{}}
        \toprule
        & PR & ESC-50 & NSynth \\
        Method & PER$\downarrow$ & Acc$\uparrow$ & Acc$\uparrow$ \\
        \midrule
        USAD~\cite{chang2025usad} & 8.8 & 80.3 & 55.1 \\
        USAD 2.0 (proposed) & 8.7 & \textbf{85.7} & \textbf{70.3} \\
        ~~~w/o Domain-aware Distillation & 13.3 & 83.4 & 69.1 \\
        ~~~w/o Music Domain Teacher & 8.5 & 85.2 & 49.1 \\
        ~~~w/o Music Data & \textbf{8.4} & 84.3 & 53.2 \\
        \bottomrule
    \end{tabular}
    \end{threeparttable}
    \end{adjustbox}
    \vspace{-8pt}
\end{table}

\begin{table}[t]
    \centering
    \caption{
        Ablation studies on the initialization approaches.
    }
    \label{tab:ablation-plus-scale}
    \vspace{-8pt}
    \begin{adjustbox}{max width=\linewidth}
    \begin{threeparttable}
    \begin{tabular}{@{}lcc@{}}
        \toprule
        & \multicolumn{2}{c@{}}{XARES-LLM} \\
        USAD 2.0+ & Track A & Track B \\
        \midrule
        XLarge+ \\
        ~~~from scratch & 0.731 & 0.574 \\
        ~~~init with XLarge~(proposed) & \textbf{0.772} & \textbf{0.611} \\
        \midrule
        \multicolumn{2}{@{}l}{XXLarge+ (init with depth-scaled XLarge)} \\
        ~~~new top 16 layers & 0.779 & 0.617 \\
        ~~~uniform layer duplication & 0.775 & 0.622 \\
        ~~~depth up-scaling~\cite{kim2024solar} & \textbf{0.783} & \textbf{0.624} \\
        \bottomrule
    \end{tabular}
    \end{threeparttable}
    \end{adjustbox}
    \vspace{-5pt}
\end{table}

This section ablates and analyzes the proposed techniques for USAD 2.0 training and scaling.
As shown in Fig.~\ref{fig:domain-scale}, the domain-aware distillation scale $\alpha$ achieves the best performance across domain tasks when set to 10.
When $\alpha$ is too small, the student learns from weaker targets due to mismatches between the expert and the data domain in the USAD approach~\cite{chang2025usad}.
Meanwhile, an overly large $\alpha$ degrades performance, indicating that excessively strong supervision from matched-domain experts can reduce the cross-domain generalizability.

We conduct ablation studies on the proposed methods in Tab.~\ref{tab:ablation}.
USAD 2.0 surpasses USAD~\cite{chang2025usad} under the same training and data setup.
Next, results without domain-aware distillation further demonstrate the importance of this technique for balancing performance across domains.
Without the music-domain teacher for distillation, the pitch classification accuracy drops by 30\%~(relative).
A similar phenomenon is observed when music data is removed, implying the necessity of both domain experts and in-domain data for music.

Furthermore, we evaluate the proposed initialization and depth-scaling approaches in Tab.~\ref{tab:ablation-plus-scale}.
First, initializing the supervised XLarge+ model from the SSL-pretrained XLarge backbone yields substantial gains on the XARES-LLM benchmark compared to training from scratch, improving Track A from 0.731 to 0.772 and Track B from 0.574 to 0.611.
Next, we investigate three methods for scaling the 32-layer model to a 48-layer XXLarge+ architecture.
Specifically, ``new top 16 layers'' appends randomly initialized transformer encoder layers on top of the original model; ``uniform layer duplication'' duplicates every even-numbered layer for a 1.5$\times$ expansion; and ``depth up-scaling'' follows \cite{kim2024solar} by copying and stacking the first 24 and last 24 layers.
These depth-scaled models are trained via domain-aware distillation from supervised teachers.
The increased capacity allows all depth-scaling variants to outperform the XLarge+ baseline, with depth up-scaling achieving the highest overall performance on both Tracks A and B.
Collectively, these ablation studies confirm the efficacy of the proposed USAD 2.0 training and scaling strategies.

\subsection{Inference Efficiency}
\label{sec:exp-inference-efficiency}

\begin{table}[t]
    \centering
    \caption{
        Inference efficiency of different model sizes.
        The metrics are measured on an A5000 GPU and averaged over 50 runs, with a 30-second audio input.
    }
    \label{tab:inference-speed}
    \vspace{-8pt}
    \begin{adjustbox}{max width=\linewidth}
    \begin{tabular}{@{}l@{~~}r@{~~}c@{~~}c@{~~}c@{}}
        \toprule
        USAD 2.0 Size & Params & Framerate & RTF$\downarrow$ & \shortstack{Peak GPU\\Memory$\downarrow$} \\
        \midrule
        Large & 336M & 50Hz & 0.0029 & 1.2GB \\
        \midrule
        XLarge~(50Hz) & 695M & 50Hz & 0.0051 & 2.2GB \\
        XLarge~(25Hz, proposed) & 695M & 25Hz & 0.0018 & 1.7GB \\
        \midrule
        XXLarge~(50Hz) & 1036M & 50Hz & 0.0077 & 3.0GB \\
        XXLarge~(25Hz, proposed) & 1036M & 25Hz & 0.0026 & 2.4GB \\
        \bottomrule
    \end{tabular}
    \end{adjustbox}
    \vspace{-8pt}
\end{table}

We compare the inference efficiency of USAD 2.0 to assess its real-world applicability.
We use the real-time factor~(RTF), defined as the ratio of inference time to input audio duration, to quantify inference speed.
As shown in Tab.~\ref{tab:inference-speed}, the Large model has the lowest memory usage because of the model size.
Furthermore, the proposed framerate reduction yields substantial computational benefits.
Specifically, the 25Hz XLarge model speeds up by more than 2.8$\times$ and reduces memory usage by 20\% compared with the 50Hz counterpart.
The 25Hz XXLarge model, despite scaling to over one billion parameters, operates faster than the 336M-parameter Large model at 50Hz, while keeping peak memory usage at a manageable 2.4GB.
Taken together with Tab.~\ref{tab:main-eval} and Tab.~\ref{tab:inference-speed}, these results show that the efficient scaling of USAD 2.0 yields continual improvements while maintaining fast inference performance.

\subsection{Representation Visualization}
\label{sec:exp-repr-visualization}

\begin{figure}[t]
    \centering
    \includegraphics[width=1.02\linewidth,trim={0.3cm 0 0 1.2cm},clip]{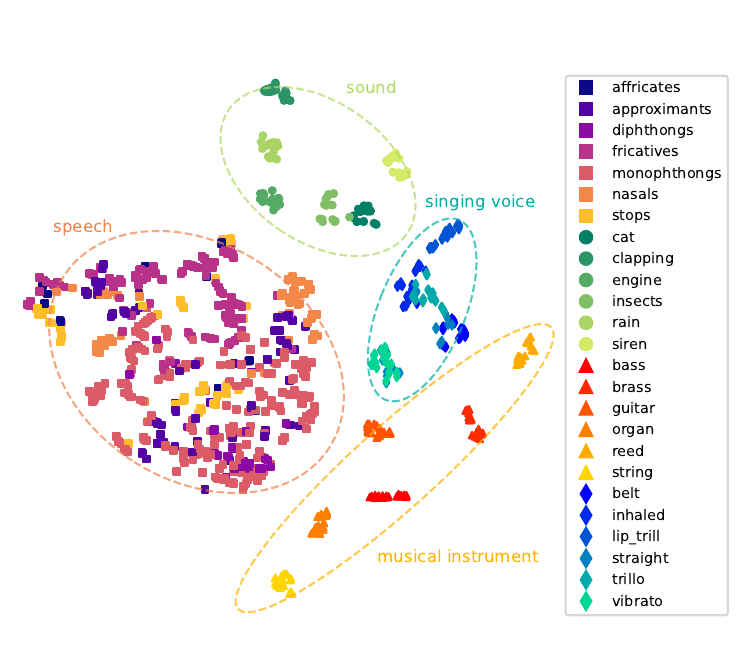}
    \vspace{-25pt}
    \caption{
        t-SNE~\cite{van2008tsne} visualization of USAD 2.0 XXLarge+ hidden representations with speech~\cite{garofolo1993timit}, environmental sounds~\cite{piczak2015esc50}, musical instruments~\cite{engel2017nsynth}, and singing voices~\cite{wilkins2018vocalset}.
    }
    \label{fig:tsne-xxl}
    \vspace{-8pt}
\end{figure}

This section visualizes the hidden representations of USAD 2.0 XXLarge+ to understand how audio is encoded into high-dimensional embedding spaces.
We visualize the 40\textsuperscript{th}-layer embeddings of USAD 2.0 XXLarge+, which achieves the best XARES-LLM performance.
The embeddings are mean-pooled along the temporal dimension for each audio clip, except for speech, where each phoneme segment is pooled.

As shown in Fig.~\ref{fig:tsne-xxl}, the embeddings form four distinct macro-clusters corresponding to the broad input domains: speech, environmental sound, singing voice, and musical instrument.
Within each domain, representations are further organized into fine-grained categories.
In particular, environmental sounds and musical instruments form tightly isolated sub-clusters, while speech phonemes exhibit a more continuous distribution with slight overlap, reflecting the connected nature of spoken articulation.
Moreover, singing voice representations also cluster by vocal techniques with some overlap.
These observations indicate the model effectively disentangles multiple input domains while preserving intra-domain categorical structure, offering well-separated representations that allow downstream models easy access to the required information.

\section{Conclusion}
\label{sec:conclusion}

This paper presents USAD 2.0, a scalable universal audio encoder for audio LLMs, combining domain-aware distillation, a music expert, and supervised distillation to integrate strengths from self-supervised and supervised foundation models.
Efficient approaches scale the model to 1B parameters within an academic budget while maintaining fast inference.
USAD 2.0 delivers strong, balanced cross-domain performance, outperforming prior universal and domain-specific encoders, making it a practical frontend for next-generation audio LLMs.

\section{Generative AI Use Disclosure}


Generative AI is used to polish the manuscript without significant changes to the authors' original draft.

\bibliographystyle{IEEEtran}
\bibliography{mybib,refs,refs_audio,references/refs_llm,references/refs_lalm,references/refs_data,references/refs_my_papers}

\clearpage
\appendix

\section{Training Setup}

\subsection{Data}

As shown in Tab.~\ref{tab:dataset}, we construct a large multi-domain audio dataset by combining publicly available multilingual speech, general audio, and music corpora.
The first-stage SSL distillation uses all datasets in the first three sections of Tab.~\ref{tab:dataset} to encourage domain diversity.
For second-stage supervised distillation, we remove several smaller and noisier datasets to stabilize training.
Since fine-grained speech representations are more difficult to distill than sound and music representations, speech accounts for roughly half of the training data.
Because the model already acquires strong music capability during the first stage, we reduce the music proportion from 15\% to 9\% in the second stage.

\subsection{USAD 2.0 Training}

The complete hyperparameters are reported in Tab.~\ref{tab:usad2-hparam}.
Following USAD~\cite{chang2025usad}, the audio waveform is first converted into 128-bin Mel spectrogram features.
The USAD~2.0 backbone comprises a two-layer CNN feature extractor, a five-layer convolutional positional encoding module~\cite{baevski2020wav2vec2}, and a transformer encoder.
For first-stage training, USAD~2.0 distills from WavLM~\cite{chen2022wavlm}, ATST-Frame~\cite{li2024atst}, and MuQ~\cite{zhu2025muq}, which serve as domain experts for speech, general audio, and music, respectively.
This stage directly evaluates the extension from the two-teacher setting of USAD to three SSL teachers.
For second-stage supervised distillation, the targets are the Whisper Large-v3 encoder~\cite{radford2022whisper} and Audio Flamingo 3~(AF3) AF-Whisper~\cite{goel2025af3}.
Only the last layer of each supervised expert is distilled, and losses from all domains are weighted equally because AF3 is multi-domain.
The domain-aware distillation scale $\alpha$ is set to 10 for all models.

\begin{table}[t]
    \centering
    \caption{
        Datasets for USAD 2.0 training.
        The dataset sizes might differ from the original ones due to preprocessing.
        $^{\spadesuit}$ indicates the datasets removed after the first-stage training.
    }
    \label{tab:dataset}
    \vspace{-8pt}
    \begin{adjustbox}{max width=\linewidth}
    \begin{tabular}{@{}l@{~~~~}r@{~~~~}r@{~~~~}r@{}}
        \toprule
        Dataset & Clips & \shortstack{Duration\\(hours)} & Proportion \\
        \midrule
        \textbf{Speech} \\
        ~~LibriVox~\cite{Panayotov2015libri,kahn2020libri,pratap2020mls} & 13,841,487 & 55,608 & 42.24\% \\
        ~~Common Voice 17~\cite{ardila2020commonvoice} & 5,455,997 & 9,023 & 16.65\% \\
        ~~VoxPopuli (English)~\cite{wang2021voxpopuli} & 3,051,826 & 24,084 & 9.31\% \\
        ~~MMS unlabeled v2~\cite{chen2024xeus,pratap2024mms}$^{\spadesuit}$ & 2,904,325 & 7,410 & 8.86\% \\
        ~~GigaSpeech~\cite{chen2021gigaspeech} & 2,568,818 & 5,306 & 7.84\% \\
        ~~VoxLingua107~\cite{valk2021voxlingua107} & 2,076,677 & 5,726 & 6.34\% \\
        ~~MLS~(non-English)~\cite{pratap2020mls} & 1,445,342 & 6,027 & 4.41\% \\
        ~~Fisher~\cite{cieri2004fisher}$^{\spadesuit}$ & 623,915 & 1,277 & 1.90\% \\
        ~~BABEL~\cite{roach1998babel} & 381,220 & 798 & 1.16\% \\
        ~~FLEURS~\cite{conneau2023fleurs} & 267,042 & 951 & 0.81\% \\
        ~~AISHELL-1~\cite{bu2017aishell} & 69,958 & 105 & 0.21\% \\
        ~~CHiME-3~\cite{barker2015chime3}$^{\spadesuit}$ & 50,180 & 111 & 0.15\% \\
        ~~Zeroth-Korean$^{\spadesuit}$ & 17,892 & 41 & 0.05\% \\
        ~~Expresso~\cite{nguyen2023expresso}$^{\spadesuit}$ & 14,468 & 31 & 0.04\% \\
        \midrule
        \textbf{General Audio} \\
        ~~SoundNet~\cite{aytar2016soundnet}$^{\spadesuit}$ & 4,637,914 & 12,385 & 57.52\% \\
        ~~AudioSet~\cite{gemmeke2017audioset} & 1,929,338 & 5,318 & 23.93\% \\
        ~~LAION-Audio-630k~\cite{wu2023laion-audio} & 1,311,739 & 3,438 & 16.27\% \\
        ~~VGGSound~\cite{chen2020vggsound} & 183,721 & 510 & 2.28\% \\
        \midrule
        \textbf{Music} \\
        ~~FMA~\cite{defferrard2016fma} & 2,961,775 & 8,173 & 58.14\% \\
        ~~MTG-Jamendo~\cite{bogdanov2019mtg} & 1,365,876 & 3,766 & 26.81\% \\
        ~~Music4All~\cite{santana2020music4all} & 327,807 & 911 & 6.44\% \\
        ~~NSynth~\cite{engel2017nsynth} & 289,205 & 321 & 5.68\% \\
        ~~MagnaTagATune~\cite{law2009mtt} & 77,580 & 209 & 1.52\% \\
        ~~MAESTRO~\cite{hawthorne2018maestro} & 71,847 & 199 & 1.41\% \\
        \midrule
        \multicolumn{3}{@{}l@{~~~~}}{\textbf{Multi-domain Dataset (Self-supervised Teachers)}} \\
        ~~Speech (1$\times$ upsample) & 32,769,147 & 116,495 & 48.80\% \\
        ~~General Audio (3$\times$ upsample) & 24,188,136 & 64,954 & 36.02\% \\
        ~~Music (2$\times$ upsample) & 10,188,180 & 27,158 & 15.17\% \\
        Total & 67,145,463 & 208,607 & 100.00\% \\
        \midrule
        \multicolumn{3}{@{}l@{~~~~}}{\textbf{Multi-domain Dataset (Supervised Teachers)}} \\
        ~~Speech (1$\times$ upsample) & 29,158,367 & 107,627 & 53.21\% \\
        ~~General Audio (6$\times$ upsample) & 20,548,788 & 55,598 & 37.50\% \\
        ~~Music (1$\times$ upsample) & 5,094,090 & 13,579 & 9.30\% \\
        Total & 54,801,245 & 176,805 & 100.00\% \\
        \bottomrule
    \end{tabular}
    \end{adjustbox}
    \vspace{-8pt}
\end{table}

\begin{table*}[t]
    \centering
    \caption{
        Hyperparameters of USAD 2.0.
    }
    \label{tab:usad2-hparam}
    \vspace{-8pt}
    \begin{adjustbox}{max width=\linewidth}
    \begin{tabular}{@{}l@{~}cccccccc@{}}
        \toprule
         & \shortstack{Small \\ Ablation} & Small & Base & Large & XLarge & Large+ & XLarge+ & XXLarge+ \\
        \midrule
        \textbf{Model} \\
        ~~Initialization & -- & -- & -- & -- & -- & \scriptsize \shortstack{USAD 2.0 \\ Large} & \scriptsize \shortstack{USAD 2.0 \\ XLarge} & \scriptsize \shortstack{USAD 2.0 \\ XLarge} \\[2pt]
        ~~Parameters & 25M & 25M & 97M & 336M & 695M & 336M & 695M & 1036M \\
        ~~Hidden Size & 384 & 384 & 768 & 1024 & 1280 & 1024 & 1280 & 1280 \\
        ~~FFN Size & 1536 & 1536 & 3072 & 4096 & 5120 & 4096 & 5120 & 5120 \\
        ~~Layers & 12 & 12 & 12 & 24 & 32 & 24 & 32 & 48 \\
        ~~Heads & 6 & 6 & 12 & 16 & 20 & 16 & 20 & 20 \\
        ~~Framerate & 50Hz & 50Hz & 50Hz & 50Hz & 25Hz & 50Hz & 25Hz & 25Hz \\
        \midrule
        \textbf{Teachers} \\
        ~~Speech & \scriptsize \shortstack{WavLM\\Base+} & \scriptsize \shortstack{WavLM\\Base+} & \scriptsize \shortstack{WavLM\\Base+} & \scriptsize \shortstack{WavLM\\Large} & \scriptsize \shortstack{WavLM\\Large} & \scriptsize \shortstack{Whisper\\Large-v3} & \scriptsize \shortstack{Whisper\\Large-v3} & \scriptsize \shortstack{Whisper\\Large-v3} \\[2pt]
        ~~Audio & \scriptsize \shortstack{ATST\\Frame} & \scriptsize \shortstack{ATST\\Frame} & \scriptsize \shortstack{ATST\\Frame} & \scriptsize \shortstack{ATST\\Frame} & \scriptsize \shortstack{ATST\\Frame} & \scriptsize \shortstack{Audio \\ Flamingo 3} & \scriptsize \shortstack{Audio \\ Flamingo 3} & \scriptsize \shortstack{Audio \\ Flamingo 3} \\[2pt]
        ~~Music & \scriptsize MuQ\textsubscript{iter} & \scriptsize MuQ\textsubscript{iter} & \scriptsize MuQ\textsubscript{iter} & \scriptsize MuQ\textsubscript{iter} & \scriptsize MuQ\textsubscript{iter} & -- & -- & -- \\
        \midrule
        \textbf{Training} \\
        ~~Optimizer & \multicolumn{8}{c}{Adam} \\
        ~~Learning Rate & 5e-4 & 8e-4 & 1.2e-3 & 1.5e-3 & 2e-3 & 5e-4 & 5e-4 & 5e-4 \\
        ~~LR Warmup & 4k & 32k & 32k & 32k & 32k & 4k & 4k & 4k \\
        ~~Updates & 150k & 600k & 600k & 600k & 600k & 50k & 50k & 50k \\
        ~~Batch Size & 200s & 800s & 800s & 1200s & 1200s & 800s & 800s & 800s \\
        ~~A6000 GPUs & 1 & 4 & 4 & 4 & 4 & 4 & 4 & 4 \\
        \bottomrule
    \end{tabular}
    \end{adjustbox}
\end{table*}

\section{Additional Results}

\subsection{Probing and Fine-tuning Benchmarks}

We provide complete experimental results for these tasks:
\begin{itemize}
    \item Audio Tagging and Sound Classification: Tab.~\ref{tab:audio-ft-eval}.
    \item HEAR~\cite{turian2022hear}: Tab.~\ref{tab:hear-eval-full}.
    \item MARBLE~\cite{yuan2023marble}: Tab.~\ref{tab:marble-eval}.
    \item SUPERB~\cite{yang2021superb}: Tab.~\ref{tab:superb-eval}.
\end{itemize}

\begin{table*}[t]
    \centering
    \caption{
        Results on audio tagging~(AS-20K)~\cite{gemmeke2017audioset} and sound classification~(ESC-50)~\cite{piczak2015esc50}.
        The audio encoders are fully fine-tuned.
    }
    \label{tab:audio-ft-eval}
    \vspace{-8pt}
    \begin{adjustbox}{max width=\linewidth}
    \begin{threeparttable}
    \begin{tabular}{@{}lrcc@{}}
        \toprule
        & & AS-20K & ESC-50 \\
        Model & Params & mAP$\uparrow$ & Acc$\uparrow$ \\
        \midrule
        \multicolumn{2}{@{}l}{\textbf{Audio SSL Models}} \\
        ~~BEATs iter3~\cite{chen2022beats} & 90M & 38.3 & 95.6 \\
        ~~EAT~\cite{chen2024eat} & 88M & 40.2 & 95.9 \\
        ~~SSLAM~\cite{alex2025sslam} & 88M & \textbf{40.9} & 96.2 \\
        ~~ATST Frame~\cite{li2024atst} & 86M & 39.0 & 91.1 \\
        \midrule
        
        \multicolumn{4}{@{}l}{\textbf{Multi-domain Audio Models}} \\
        ~~USAD Small~\cite{chang2025usad} & 25M & 34.5 & 89.3 \\
        ~~USAD Base~\cite{chang2025usad} & 97M & 35.7 & 91.1 \\
        ~~USAD Large~\cite{chang2025usad} & 336M & 37.4 & 92.7 \\
        ~~SPEAR Base~\cite{yang2025spear} & 94M & 39.1 & -- \\
        ~~SPEAR Large~\cite{yang2025spear} & 327M & 39.2 & -- \\
        ~~SPEAR XLarge~\cite{yang2025spear} & 600M & 39.4 & -- \\
        \midrule
        
        \multicolumn{4}{@{}l}{\textbf{USAD 2.0 (Self-supervised Teachers)}} \\
        ~~USAD 2.0 Small & 25M & 37.2 & 93.5 \\
        ~~USAD 2.0 Base & 97M & 40.0 & 94.6 \\
        ~~USAD 2.0 Large & 336M & 40.1 & 95.0 \\
        \midrule
        
        \multicolumn{4}{@{}l}{\textbf{USAD 2.0+ (Supervised Teachers)}} \\
        ~~USAD 2.0 Large+ & 336M & 40.6 & \textbf{96.8} \\
        ~~USAD 2.0 XLarge+ & 695M & \textbf{40.9} & 96.4 \\
        \bottomrule
    \end{tabular}
    \end{threeparttable}
    \end{adjustbox}
\end{table*}

\begin{table*}[t]
    \centering
    \caption{
        Results on HEAR~\cite{turian2022hear}.
    }
    \label{tab:hear-eval-full}
    \vspace{-8pt}
    \begin{adjustbox}{max width=\linewidth}
    \begin{threeparttable}
    \begin{tabular}{@{}l@{~~}r@{~~}c@{~~}c@{~~}c@{~~}c@{~~}c@{~~}c@{~~}c@{~~}c@{~~}c@{~~}c@{~~}c@{~~}c@{~~}c@{~~}c@{~~}c@{~~}c@{~~}c@{~~}c@{~~}c@{~~}c@{~~}c@{~~}c@{}}
        \toprule
         & & & & & & & & & & & & & & & & & & & & \multicolumn{4}{@{~~}c@{~~}}{\scriptsize Average} \\
        \cmidrule{21-24}
        Model & Params & \scriptsize BJ & \scriptsize CD & \scriptsize D16 & \scriptsize E50 & \scriptsize F50k & \scriptsize Gun & \scriptsize GZ-Gen & \scriptsize GZ-M/S & \scriptsize LiCt & \scriptsize MST & \scriptsize Mri-S & \scriptsize Mri-T & \scriptsize SPC-5 & \scriptsize SPC-F & \scriptsize NS-5 & \scriptsize NS-50 & \scriptsize VI & \scriptsize VL & \scriptsize Speech & \scriptsize Env & \scriptsize Music & \scriptsize All \\
        \midrule
        \multicolumn{2}{@{}l@{~~}}{\textbf{Audio Models}} \\
        ~~ATST Frame~\cite{li2024atst} & 86M & 95.8 & 76.7 & 95.7 & 89.0 & 55.7 & 94.3 & 88.3 & 100.0 & 78.1 & 24.4 & 97.5 & 96.9 & 92.6 & 95.1 & 68.6 & 82.0 & 22.3 & 66.9 & 72.0 & 83.7 & 81.7 & 78.9 \\
        ~~Dasheng Base~\cite{dinkel24bdasheng} & 86M & 93.6 & 78.7 & 95.7 & 82.9 & 51.0 & 92.9 & 89.2 & 99.2 & 76.6 & 43.9 & 96.1 & 94.9 & 95.9 & 97.1 & 71.8 & 83.3 & 16.7 & 69.9 & 72.5 & 80.6 & 84.0 & 79.4 \\
        ~~Dasheng 0.6B~\cite{dinkel24bdasheng} & 630M & 94.9 & 81.2 & 94.4 & 85.9 & 53.9 & 97.6 & 88.6 & 97.6 & 80.7 & 43.5 & 96.6 & 96.2 & 97.0 & 97.5 & 74.6 & 85.8 & 17.8 & 74.7 & 74.8 & 83.0 & 84.7 & 81.0 \\
        ~~Dasheng 1.2B~\cite{dinkel24bdasheng} & 1.2B & 96.2 & 81.6 & 94.2 & 85.3 & 54.2 & 99.1 & 88.8 & 99.1 & 79.6 & 43.3 & 96.8 & 96.1 & 97.1 & 97.9 & 74.4 & 85.6 & 19.4 & 78.7 & 75.7 & 83.2 & 84.9 & 81.4 \\
        \midrule
        \multicolumn{5}{@{}l@{~~}}{\textbf{Multi-domain Audio Models}} \\
        ~~USAD Small~\cite{chang2025usad} & 25M & 94.5 & 78.2 & 89.5 & 81.8 & 51.1 & 93.2 & 86.6 & 98.5 & 77.0 & 25.3 & 97.3 & 94.3 & 96.2 & 97.2 & 55.6 & 77.7 & 20.0 & 73.6 & 73.7 & 78.9 & 78.7 & 77.1 \\
        ~~USAD Base~\cite{chang2025usad} & 97M & 95.8 & 80.0 & 93.6 & 82.2 & 52.2 & 94.0 & 86.3 & 100.0 & 78.7 & 26.7 & 97.3 & 95.7 & 96.6 & 97.6 & 57.0 & 81.6 & 19.5 & 76.0 & 74.7 & 80.5 & 80.0 & 78.4 \\
        ~~USAD Large~\cite{chang2025usad} & 336M & 94.1 & 79.5 & 93.9 & 83.4 & 53.0 & 97.6 & 87.4 & 100.0 & 79.1 & 38.4 & 97.4 & 96.1 & 97.0 & 97.5 & 57.0 & 83.2 & 18.5 & 75.3 & 74.5 & 82.0 & 81.7 & 79.4 \\
        ~~SPEAR Base~\cite{yang2025spear} & 94M & 95.3 & 82.0 & 95.1 & 85.9 & 54.2 & 95.2 & 88.8 & 100.0 & 76.2 & 26.8 & 97.2 & 96.0 & 97.3 & 98.2 & 69.4 & 82.2 & 24.6 & 85.6 & 77.3 & 82.6 & 82.0 & 80.6 \\
        ~~SPEAR Large~\cite{yang2025spear} & 327M & 94.9 & 83.8 & 95.9 & 87.6 & 56.4 & 97.6 & 89.2 & 99.2 & 78.7 & 27.9 & 97.4 & 97.5 & 98.1 & 98.3 & 70.2 & 85.3 & 25.7 & 88.5 & 78.9 & 84.4 & 82.7 & 81.8 \\
        ~~SPEAR XLarge~\cite{yang2025spear} & 600M & 95.3 & 83.6 & 96.0 & 89.4 & 57.1 & 96.3 & 91.0 & 100.0 & 80.7 & 27.7 & 97.4 & 97.9 & 98.4 & 98.6 & 74.2 & 86.0 & 26.6 & 90.4 & 79.7 & 84.7 & 83.7 & 82.6 \\
        ~~Whisper Large-v3~\cite{radford2022whisper} & 637M & 95.3 & 81.8 & 93.4 & 86.1 & 42.4 & 92.0 & 87.6 & 96.0 & 70.6 & 24.5 & 97.4 & 94.9 & 98.5 & 98.6 & 62.4 & 66.7 & 25.6 & 98.5 & 78.9 & 78.5 & 78.1 & 78.5 \\
        ~~AF3 Whisper~\cite{goel2025af3} & 637M & 96.2 & 85.5 & 93.8 & 94.3 & 61.4 & 97.6 & 94.8 & 96.9 & 71.8 & 28.7 & 97.8 & 97.1 & 97.4 & 98.2 & 71.4 & 81.8 & 25.7 & 92.9 & 78.6 & 86.8 & 83.1 & 82.4 \\
        \midrule

        \multicolumn{5}{@{}l@{~~}}{\textbf{Multi-expert Encoder}} \\
        ~~Self-supervised \\
        ~~~{\scriptsize (WavLM + ATST + MuQ)} & 734M & 95.8 & 80.3 & 93.6 & 83.4 & 54.5 & 97.6 & 86.8 & 97.7 & 79.7 & 42.4 & 97.7 & 98.1 & 96.6 & 97.7 & 75.8 & 88.7 & 24.6 & 84.1 & 77.2 & 82.3 & 85.4 & 82.0 \\
        ~~Supervised \\
        ~~~{\scriptsize (Whisper + AF3)} & 1274M & 95.3 & 85.2 & 93.4 & 92.3 & 54.7 & 95.2 & 92.8 & 96.1 & 70.6 & 27.9 & 97.8 & 96.1 & 98.1 & 98.4 & 71.4 & 81.8 & 27.1 & 98.4 & 79.6 & 83.9 & 82.4 & 81.8 \\
        \midrule
        
        \multicolumn{5}{@{}l@{~~}}{\textbf{USAD 2.0}} \\
        ~~USAD 2.0 Small & 25M & 95.8 & 77.8 & 94.3 & 85.3 & 53.7 & 92.9 & 87.3 & 100.0 & 78.2 & 46.4 & 97.3 & 97.5 & 96.3 & 97.0 & 76.8 & 87.4 & 21.5 & 72.3 & 73.9 & 81.5 & 86.1 & 81.0 \\
        ~~USAD 2.0 Base & 97M & 95.3 & 78.7 & 95.1 & 86.6 & 55.7 & 92.3 & 89.9 & 100.0 & 80.1 & 49.4 & 97.7 & 97.9 & 96.1 & 96.7 & 79.2 & 89.3 & 22.2 & 72.2 & 74.3 & 82.4 & 87.3 & 81.9 \\
        ~~USAD 2.0 Large & 336M & 96.2 & 79.1 & 95.0 & 87.7 & 56.0 & 97.6 & 91.1 & 100.0 & 80.6 & 50.5 & 97.9 & 98.3 & 96.4 & 97.1 & 80.6 & 89.9 & 24.1 & 73.6 & 75.1 & 84.1 & 88.1 & 82.9 \\
        ~~USAD 2.0 XLarge & 695M & 94.9 & 79.9 & 96.2 & 88.8 & 57.1 & 95.2 & 89.5 & 100.0 & 79.4 & 48.9 & 97.4 & 98.4 & 95.6 & 97.1 & 80.4 & 90.4 & 24.4 & 71.6 & 74.7 & 84.4 & 87.5 & 82.5 \\
        \midrule

        \multicolumn{5}{@{}l@{~~}}{\textbf{USAD 2.0+}} \\
        ~~USAD 2.0 Large+ & 336M & 96.2 & 80.8 & 95.3 & 93.1 & 62.2 & 96.7 & 91.4 & 100.0 & 79.6 & 43.4 & 97.9 & 97.6 & 97.8 & 98.1 & 73.6 & 87.7 & 26.1 & 95.0 & 79.5 & 86.8 & 86.0 & 84.0 \\
        ~~USAD 2.0 XLarge+ & 695M & 96.6 & 81.9 & 95.3 & 93.8 & 62.8 & 95.2 & 91.9 & 100.0 & 79.4 & 43.4 & 97.7 & 98.0 & 97.8 & 98.0 & 76.4 & 88.2 & 26.2 & 96.0 & 79.9 & 86.8 & 86.5 & 84.4 \\
        ~~USAD 2.0 XXLarge+ & 1036M & 96.2 & 82.6 & 95.0 & 93.8 & 62.5 & 96.4 & 93.2 & 100.0 & 79.7 & 42.8 & 97.5 & 97.9 & 97.5 & 97.8 & 75.6 & 88.1 & 25.8 & 96.3 & 79.9 & 86.9 & 86.4 & 84.4 \\
        \bottomrule
    \end{tabular}
    \end{threeparttable}
    \end{adjustbox}
    \vspace{-5pt}
\end{table*}

\begin{table*}[t]
    \centering
    \caption{
        Results on MARBLE~\cite{yuan2023marble}.
    }
    \label{tab:marble-eval}
    \vspace{-8pt}
    \begin{adjustbox}{max width=\linewidth}
    \begin{threeparttable}
    \begin{tabular}{@{}l@{~~}r@{}c@{~~}c@{~~}c@{~~}c@{~~}c@{~~}c@{~~}c@{~~}c@{~~}c@{~~}c@{~~}c@{~~}c@{~~}c@{~~}c@{~~}c@{}}
        \toprule
        & & GTZAN & GS & & \multicolumn{2}{c}{EMO} & & \multicolumn{2}{c}{VocalSet} & & \multicolumn{2}{c}{MTT} & & \multicolumn{2}{c}{NSynth} \\
        & & Genre & Key & & \multicolumn{2}{c}{Emotion} & & Singer & Tech & & \multicolumn{2}{c}{Tagging} & & Instr. & Pitch \\
        \cmidrule{6-7}
        \cmidrule{9-10}
        \cmidrule{12-13}
        \cmidrule{15-16}
        Model & Params & Acc$\uparrow$ & Acc\textsuperscript{R}$\uparrow$ & & R2\textsuperscript{V}$\uparrow$ & R2\textsuperscript{A}$\uparrow$ & & Acc$\uparrow$ & Acc$\uparrow$ & & ROC$\uparrow$ & AP$\uparrow$ & & Acc$\uparrow$ & Acc$\uparrow$ & Avg \\
        \midrule
        \multicolumn{2}{@{}l@{~~}}{\textbf{Music Models}} \\
        ~~MERT-330M~\cite{li2023mert} & 330M & 78.6 & 65.6 & & 61.2 & 74.7 & & 87.1 & 76.9 & & 91.3 & 40.2 & & 72.6 & 94.4 & 74.3 \\
        ~~MusicFM~\cite{won2024musicfm} & 330M & 83.8 & 63.9 & & 60.3 & 76.3 & & 92.0 & 78.4 & & 91.3 & 40.0 & & 76.2 & 91.1 & 75.3 \\
        ~~MuQ~\cite{zhu2025muq} & 330M & 85.6 & 65.0 & & 62.8 & 76.1 & & 96.2 & 81.6 & & 91.4 & 40.1 & & 79.7 & 91.2 & 77.0 \\
        
        \midrule
        \multicolumn{5}{@{}l@{~~}}{\textbf{Multi-domain Audio Models}} \\
        ~~USAD Small~\cite{chang2025usad} & 25M & 74.1 & 15.4 & & 48.1 & 71.8 & & 83.5 & 74.8 & & 91.0 & 39.6 & & 73.5 & 84.8 & 65.7 \\
        ~~USAD Base~\cite{chang2025usad} & 97M & 73.4 & 18.7 & & 44.3 & 71.9 & & 85.8 & 72.8 & & 91.2 & 40.0 & & 75.2 & 88.0 & 66.1 \\
        ~~USAD Large~\cite{chang2025usad} & 336M & 77.6 & 29.4 & & 52.6 & 75.2 & & 85.0 & 74.2 & & 91.4 & 40.6 & & 74.9 & 87.7 & 68.9 \\
        ~~SPEAR Base~\cite{yang2025spear} & 94M & 82.1 & 53.2 & & 55.7 & 75.6 & & 83.8 & 77.3 & & 91.6 & 40.7 & & 74.2 & 88.2 & 72.2 \\
        ~~SPEAR Large~\cite{yang2025spear} & 327M & 83.1 & 55.3 & & 62.5 & 76.8 & & 88.5 & 78.3 & & 91.8 & 41.3 & & 76.0 & 89.7 & 74.3 \\
        ~~SPEAR XLarge~\cite{yang2025spear} & 600M & 85.5 & 56.4 & & 62.9 & 77.9 & & 90.8 & 78.2 & & 91.8 & 41.4 & & 77.1 & 88.8 & 75.1 \\
        \midrule
        
        \multicolumn{5}{@{}l@{~~}}{\textbf{Multi-expert Encoder}} \\
        ~~Self-supervised~{\scriptsize (WavLM + ATST + MuQ)} & 734M & 85.9 & 63.5 & & 62.8 & 76.6 & & 90.2 & 79.6 & & 91.5 & 40.7 & & 79.2 & 91.2 & 76.1 \\
        ~~Supervised~{\scriptsize (Whisper + AF3)} & 1274M & 91.4 & 43.7 & & 59.9 & 77.2 & & 78.7 & 76.4 & & 92.1 & 42.6 & & 79.6 & 82.8 & 72.4 \\
        \midrule
        
        \multicolumn{5}{@{}l@{~~}}{\textbf{USAD 2.0}} \\
        ~~USAD 2.0 Small & 25M & 73.4 & 60.2 & & 59.1 & 78.0 & & 84.3 & 76.0 & & 91.5 & 40.7 & & 75.9 & 90.2 & 72.9 \\
        ~~USAD 2.0 Base & 97M & 73.8 & 62.5 & & 58.8 & 78.8 & & 88.4 & 77.5 & & 91.7 & 41.1 & & 76.5 & 91.8 & 74.1 \\
        ~~USAD 2.0 Large & 336M & 83.1 & 62.8 & & 62.2 & 78.7 & & 89.8 & 78.5 & & 91.7 & 41.3 & & 78.7 & 91.0 & 75.8 \\
        ~~USAD 2.0 XLarge & 695M & 83.4 & 62.2 & & 60.5 & 80.6 & & 89.8 & 78.4 & & 91.6 & 41.1 & & 77.7 & 91.5 & 75.7 \\
        \midrule

        \multicolumn{5}{@{}l@{~~}}{\textbf{USAD 2.0+}} \\
        ~~USAD 2.0 Large+ & 336M & 86.2 & 56.8 & & 60.9 & 78.3 & & 90.4 & 78.3 & & 91.9 & 41.6 & & 76.8 & 89.5 & 75.1 \\
        ~~USAD 2.0 XLarge+ & 695M & 86.6 & 55.1 & & 62.0 & 78.2 & & 90.3 & 78.3 & & 91.9 & 41.7 & & 76.4 & 89.9 & 75.0 \\
        ~~USAD 2.0 XXLarge+ & 1036M & 87.2 & 57.9 & & 59.7 & 80.4 & & 90.1 & 79.6 & & 91.8 & 41.4 & & 77.7 & 90.4 & 75.6 \\
        \bottomrule
    \end{tabular}
    \end{threeparttable}
    \end{adjustbox}
\end{table*}

\begin{table*}[t]
    \centering
    \caption{
        Results on SUPERB~\cite{yang2021superb}.
    }
    \label{tab:superb-eval}
    \vspace{-8pt}
    \begin{adjustbox}{max width=\linewidth}
    \begin{threeparttable}
    \begin{tabular}{@{}lrcccccccc@{}}
        \toprule
        & & \multicolumn{3}{c}{Frame-level} & & \multicolumn{4}{c}{Instance-level} \\
        \cmidrule{3-5}
        \cmidrule{7-10}
        & & PR & ASR & SD & & KS & IC & SID & ER \\
        Model & Params & PER$\downarrow$ & WER$\downarrow$ & DER$\downarrow$ & & Acc$\uparrow$ & Acc$\uparrow$ & Acc$\uparrow$ & Acc$\uparrow$ \\
        \midrule
        \multicolumn{4}{@{}l}{\textbf{Speech Models}} \\
        ~~WavLM Base+~\cite{chen2022wavlm} & 95M & 3.9 & 5.6 & 3.5 & & 97.4 & 99.0 & 89.4 & 68.7 \\
        ~~WavLM Large~\cite{chen2022wavlm} & 317M & 3.1 & 3.4 & 3.2 & & 97.9 & 99.3 & 95.5 & 70.6 \\
        \midrule
        
        \multicolumn{4}{@{}l}{\textbf{General Audio SSL}} \\
        ~~BEATs iter3~\cite{chen2022beats} & 90M & 36.4 & 25.9 & 5.2 & & 97.7 & 53.4 & 57.1 & 64.5 \\
        ~~EAT~\cite{chen2024eat} & 88M & 55.0 & 25.9 & 4.7 & & 92.8 & 53.7 & 45.0 & 62.5 \\
        ~~SSLAM~\cite{alex2025sslam} & 88M & 56.4 & 27.8 & 4.6 & & 98.8 & 51.6 & 42.6 & 62.6 \\
        ~~ATST Frame~\cite{li2024atst} & 86M & 20.4 & 18.8 & 4.7 & & 95.1 & 85.4 & 69.8 & 64.4 \\
        \midrule

        \multicolumn{4}{@{}l}{\textbf{Music SSL}} \\
        ~~MuQ~\cite{zhu2025muq} & 330M & 39.9 & 29.7 & 5.1 & & 91.5 & 57.4 & 49.7 & 62.3 \\
        \midrule
        
        \multicolumn{5}{@{}l}{\textbf{Multi-domain Audio Models}} \\
        ~~USAD Small~\cite{chang2025usad} & 25M & 7.8 & 9.5 & 4.9 & & 96.8 & 95.5 & 73.5 & 66.3 \\
        ~~USAD Base~\cite{chang2025usad} & 97M & 5.1 & 7.7 & 4.2 & & 97.1 & 98.3 & 88.6 & 68.0 \\
        ~~USAD Large~\cite{chang2025usad} & 336M & 4.0 & 6.5 & 3.9 & & 97.1 & 98.4 & 91.2 & 68.4 \\
        ~~SPEAR Base~\cite{yang2025spear} & 94M & 3.9 & 3.8 & 4.1 & & 97.6 & 98.1 & 90.0 & 69.4 \\
        ~~SPEAR Large~\cite{yang2025spear} & 327M & 3.1 & 3.4 & 3.8 & & 97.9 & 99.4 & 95.0 & 71.6 \\
        ~~SPEAR XLarge~\cite{yang2025spear} & 600M & 2.9 & 3.2 & 3.2 & & 98.1 & 99.6 & 96.3 & 73.3 \\
        \midrule
        
        \multicolumn{5}{@{}l}{\textbf{USAD 2.0 (Self-supervised Teachers)}} \\
        ~~USAD 2.0 Small & 25M & 7.5 & 9.7 & 5.4 & & 96.9 & 95.6 & 72.5 & 67.1 \\
        ~~USAD 2.0 Base & 97M & 5.3 & 8.3 & 4.8 & & 96.7 & 96.7 & 83.6 & 68.2 \\
        ~~USAD 2.0 Large & 336M & 3.9 & 5.5 & 4.8 & & 96.9 & 95.1 & 84.7 & 67.8 \\
        ~~USAD 2.0 XLarge & 695M & 3.6 & 5.6 & 4.3 & & 96.2 & 93.6 & 90.3 & 69.1 \\
        \midrule

        \multicolumn{5}{@{}l}{\textbf{USAD 2.0+ (Supervised Teachers)}} \\
        ~~USAD 2.0 Large+ & 336M & 4.3 & 5.1 & 3.6 & & 97.7 & 99.1 & 90.1 & 71.7 \\
        ~~USAD 2.0 XLarge+ & 695M & 4.1 & 5.7 & 3.6 & & 97.6 & 99.0 & 91.6 & 72.5 \\
        \bottomrule
    \end{tabular}
    \end{threeparttable}
    \end{adjustbox}
\end{table*}

\subsection{XARES-LLM Benchmark}
\begin{figure*}[t]
    \centering
    \includegraphics[width=\linewidth]{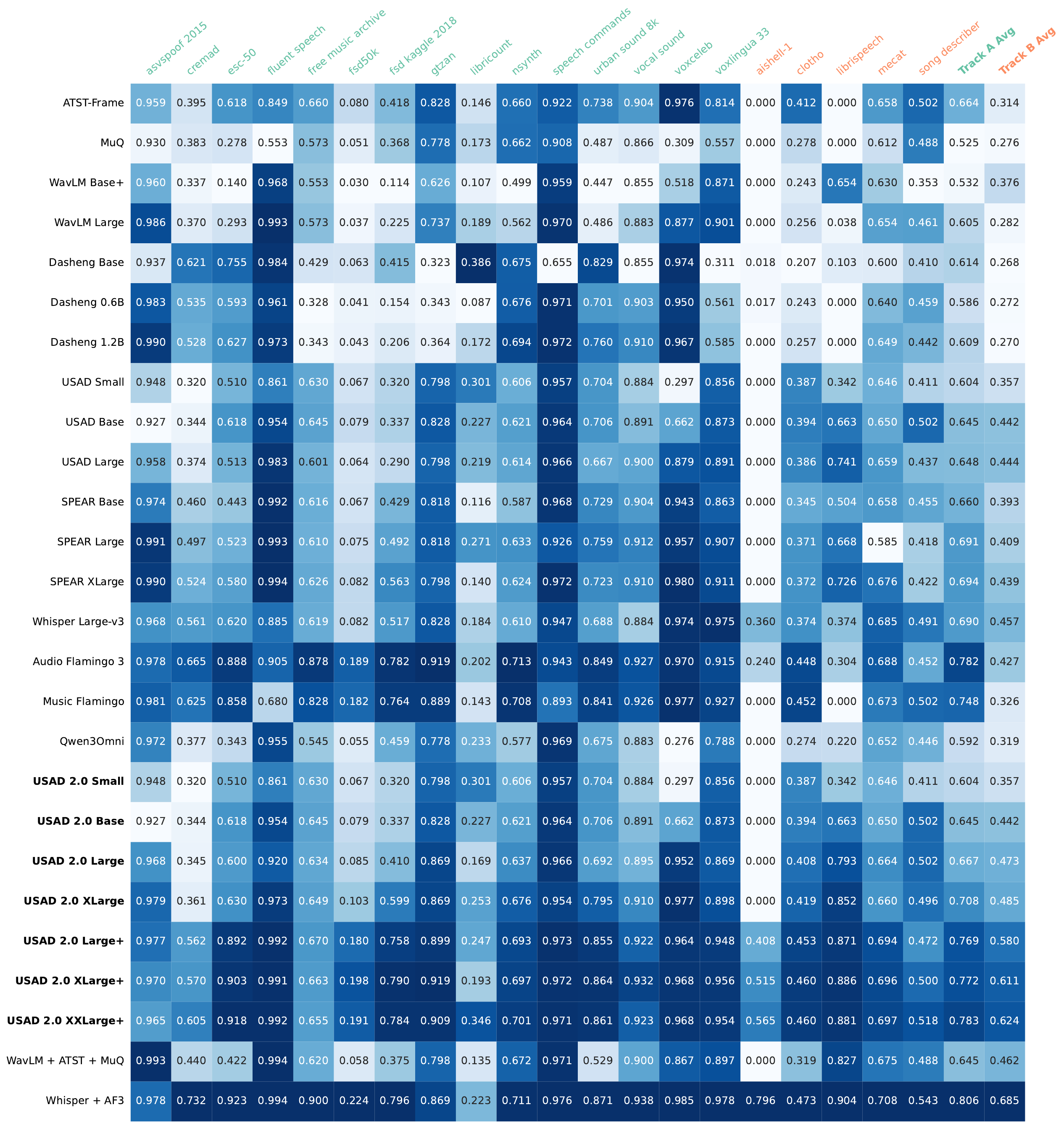}
    \caption{
        XARES-LLM results on the best-performing audio encoders.
        Columns 1--15 and 16--20 belong to Tracks A and B, respectively.
        The last two columns indicate the average scores of Track A and B, respectively.
        The colors are normalized along each column.
    }
    \label{fig:xares-llm-result}
\end{figure*}

Following the XARES-LLM benchmark~\cite{dinkel2026interspeech}, we evaluate several state-of-the-art audio encoders across different domains and report in Fig.~\ref{fig:xares-llm-result}.
Excluding multi-expert encoders, USAD 2.0 XXLarge+ is the best-performing encoder on both Track A and Track B.
Scaling from 0.3B~(Large+) to 1B~(XXLarge+) parameters yields consistent improvements across most tasks.
Comparing SSL-based and supervised encoders shows that supervised encoders are generally stronger for audio LLM applications, supporting the motivation for second-stage supervised distillation in Sec.~\ref{sec:method-scaling-supervised}.
Moreover, domain-specific encoders show strong in-domain capabilities but weaker out-of-domain performance.
For example, WavLM Large~\cite{chen2022wavlm} performs well on several speech-related tasks, whereas the Music Flamingo encoder~\cite{ghosh2025mf} largely loses speech processing ability, especially for ASR~(AISHELL-1 and LibriSpeech), after training with more music data.
In contrast, USAD~2.0 XXLarge+ exhibits more balanced performance across domains and tasks.
Overall, these results highlight the effectiveness of the proposed distillation framework and the usefulness of USAD~2.0 as a universal encoder for audio LLM applications.

\subsection{Representation Visualization}

\begin{figure*}[t]
    \centering
    \begin{subfigure}[b]{0.32\linewidth}
         \centering
         \includegraphics[width=\linewidth]{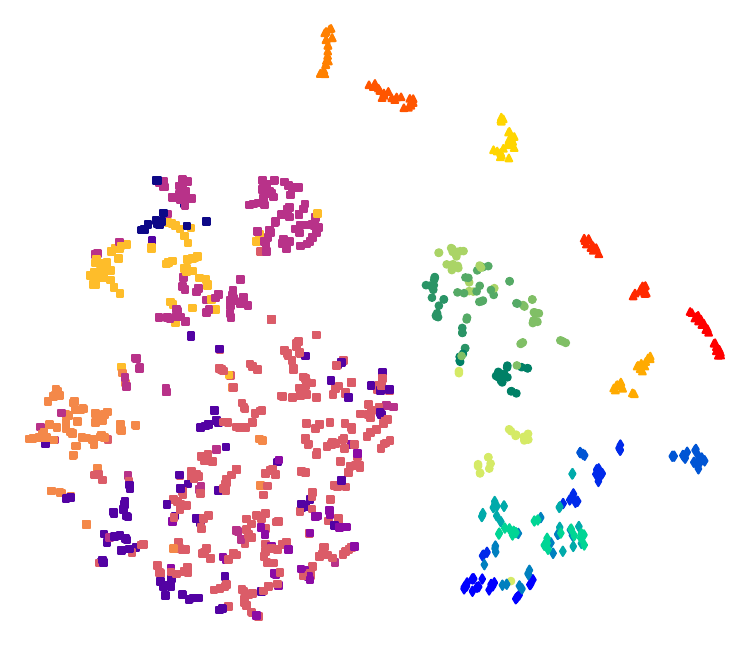}
         \caption{Layer 4}
    \end{subfigure}
    \hfill
    \begin{subfigure}[b]{0.32\linewidth}
         \centering
         \includegraphics[width=\linewidth]{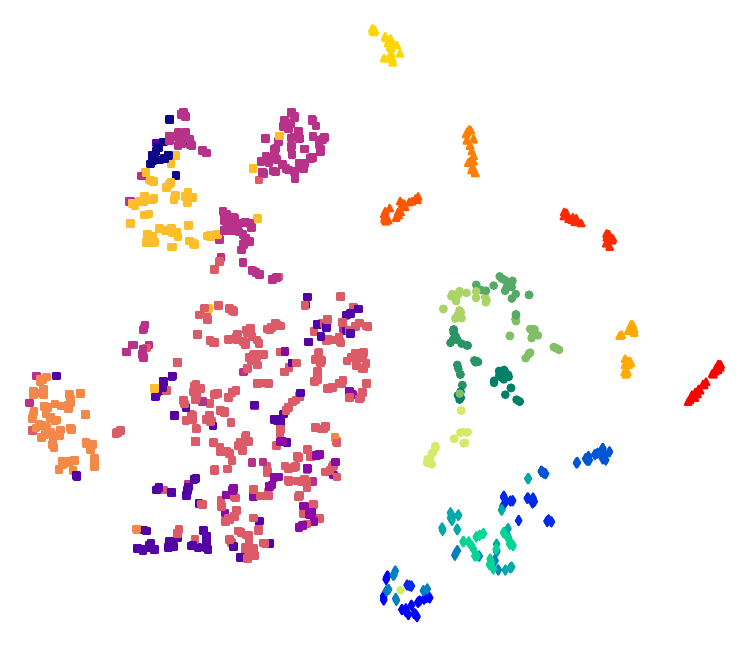}
         \caption{Layer 8}
    \end{subfigure}
    \hfill
    \begin{subfigure}[b]{0.32\linewidth}
         \centering
         \includegraphics[width=\linewidth]{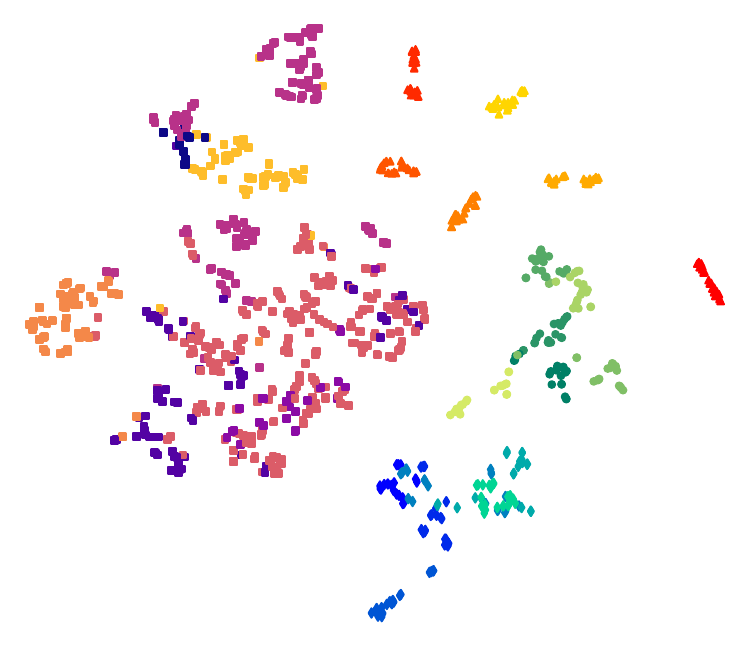}
         \caption{Layer 12}
    \end{subfigure}
    
    ~\\
    
    \begin{subfigure}[b]{0.32\linewidth}
         \centering
         \includegraphics[width=\linewidth]{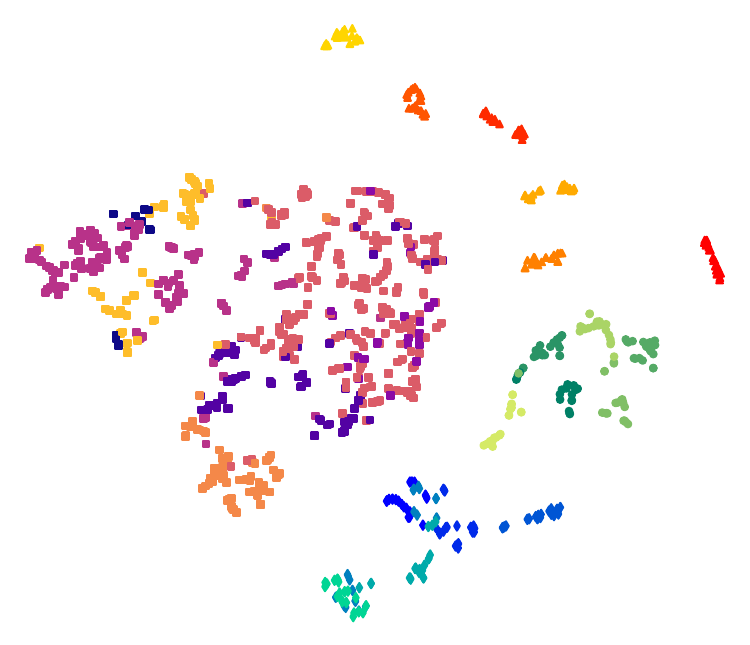}
         \caption{Layer 16}
    \end{subfigure}
    \hfill
    \begin{subfigure}[b]{0.32\linewidth}
         \centering
         \includegraphics[width=\linewidth]{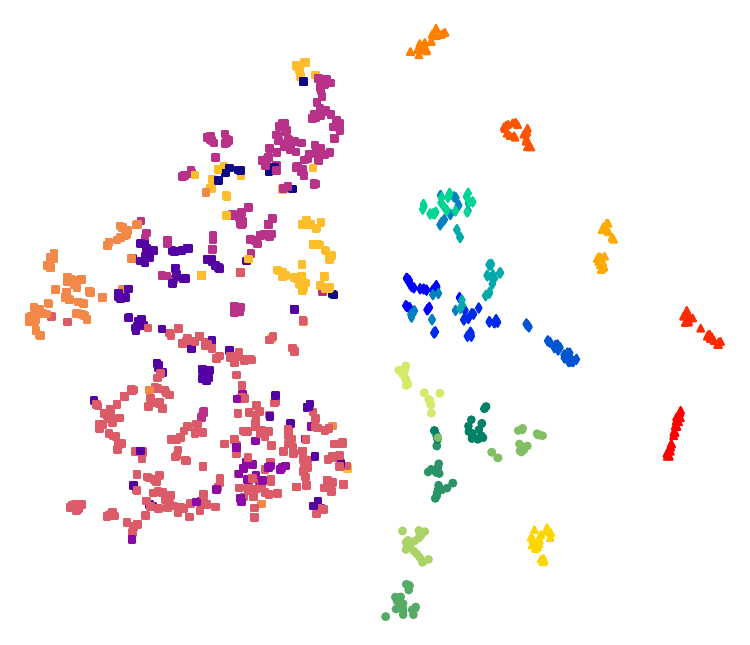}
         \caption{Layer 20}
    \end{subfigure}
    \hfill
    \begin{subfigure}[b]{0.32\linewidth}
         \centering
         \includegraphics[width=\linewidth]{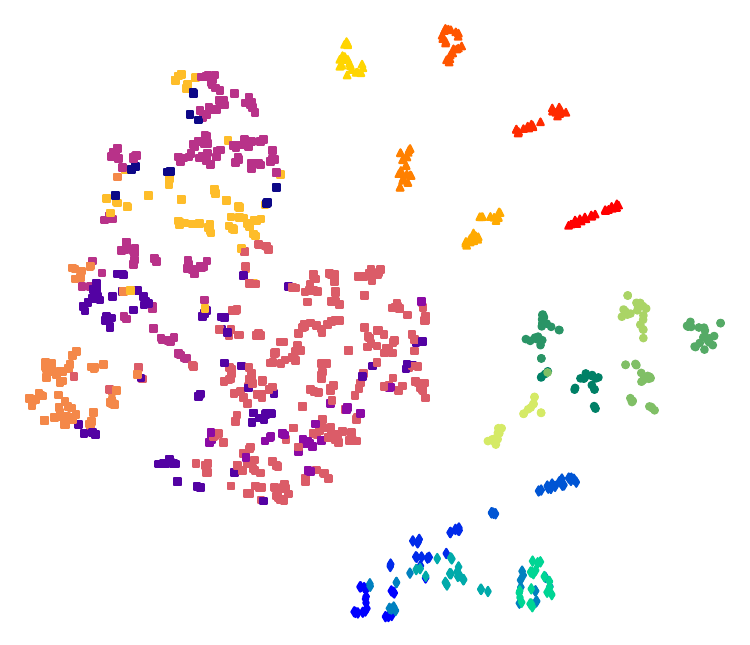}
         \caption{Layer 24}
    \end{subfigure}

    ~\\
    
    \begin{subfigure}[b]{0.32\linewidth}
         \centering
         \includegraphics[width=\linewidth]{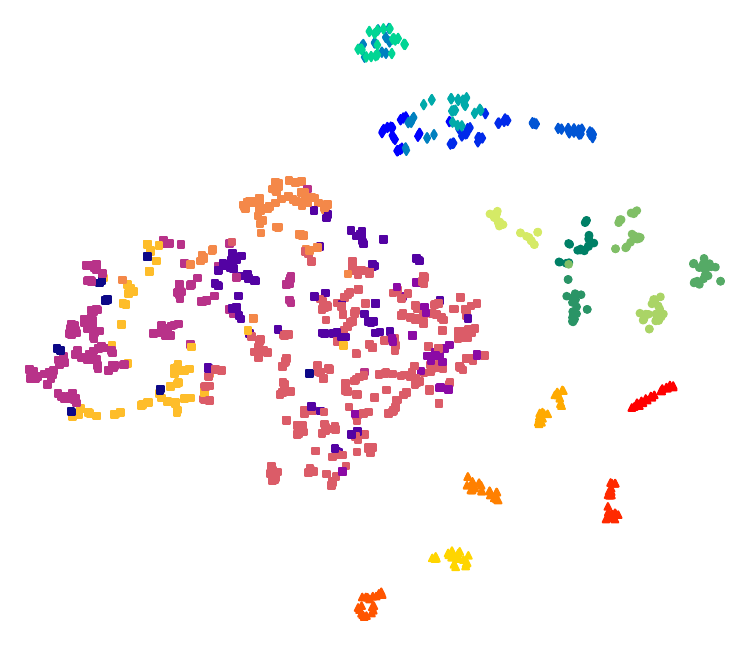}
         \caption{Layer 28}
    \end{subfigure}
    \hfill
    \begin{subfigure}[b]{0.32\linewidth}
         \centering
         \includegraphics[width=\linewidth]{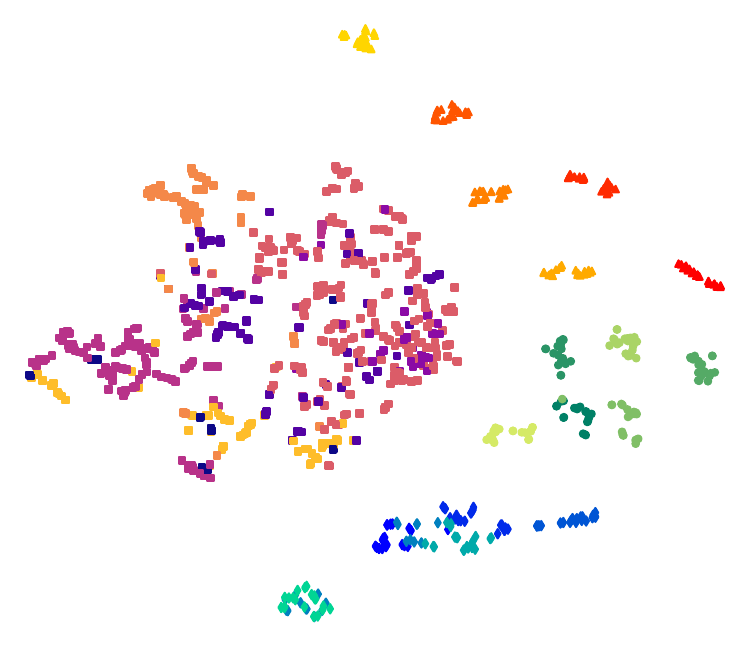}
         \caption{Layer 32}
    \end{subfigure}
    \hfill
    \begin{subfigure}[b]{0.32\linewidth}
         \centering
         \includegraphics[width=\linewidth]{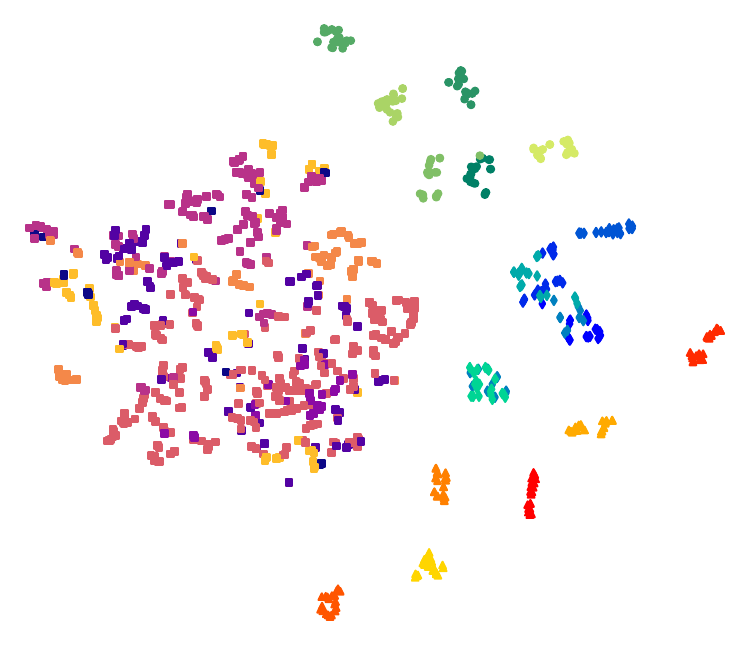}
         \caption{Layer 36}
    \end{subfigure}

    ~\\
    
    \begin{subfigure}[b]{0.32\linewidth}
         \centering
         \includegraphics[width=\linewidth]{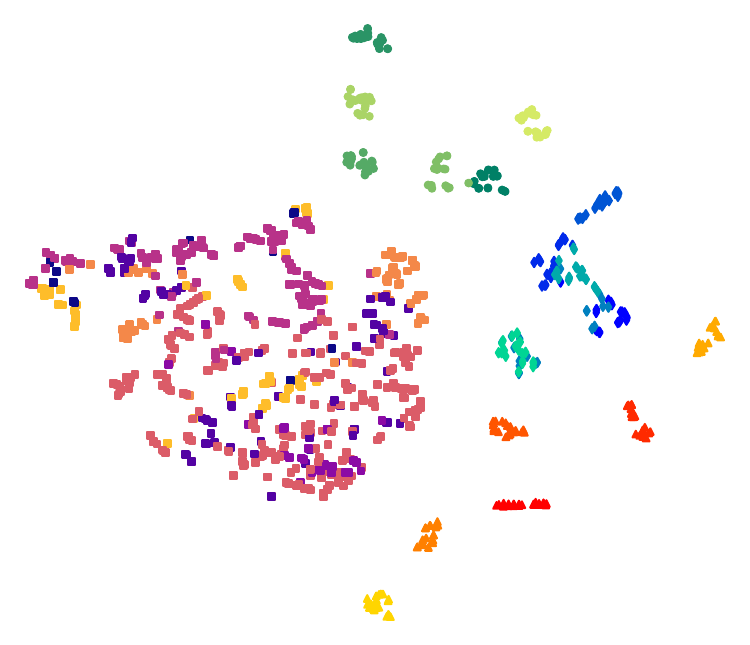}
         \caption{Layer 40}
    \end{subfigure}
    \hfill
    \begin{subfigure}[b]{0.32\linewidth}
         \centering
         \includegraphics[width=\linewidth]{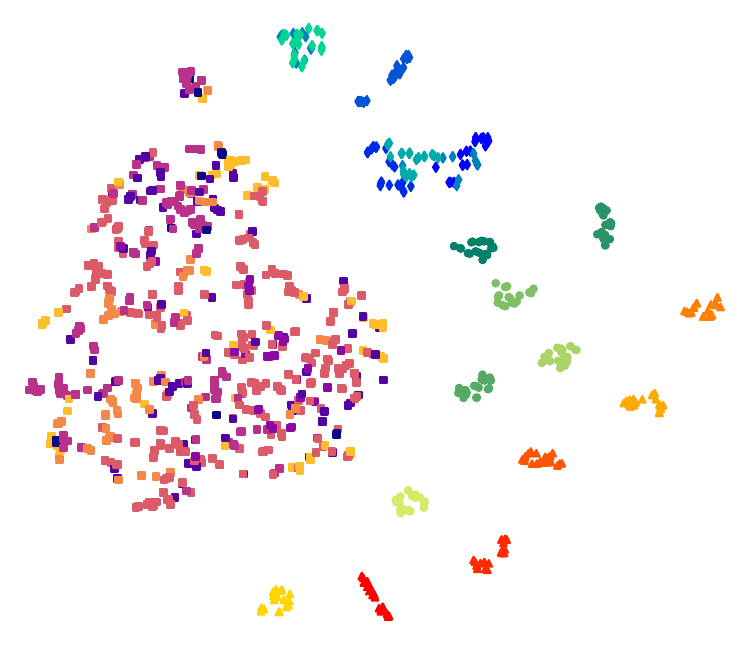}
         \caption{Layer 44}
    \end{subfigure}
    \hfill
    \begin{subfigure}[b]{0.32\linewidth}
         \centering
         \includegraphics[width=\linewidth]{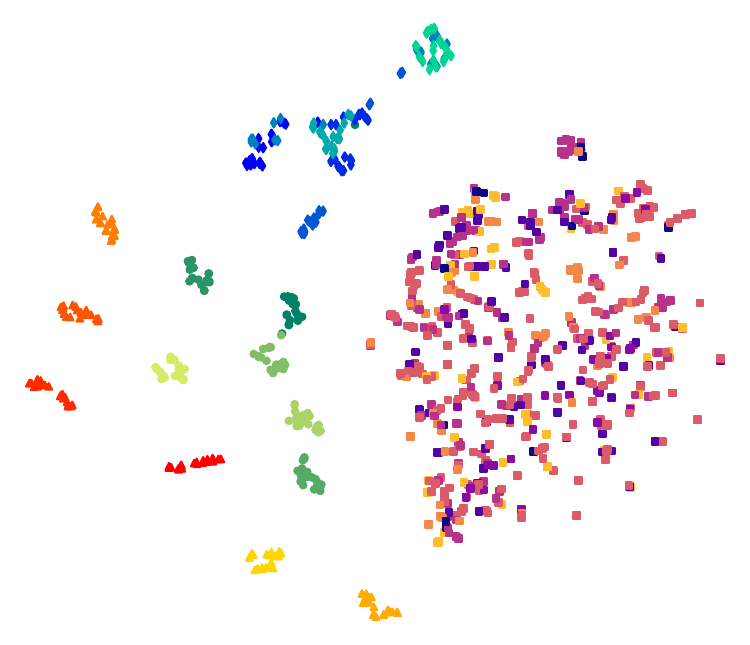}
         \caption{Layer 48}
    \end{subfigure}
     
    \caption{
        t-SNE visualization of USAD 2.0 XXLarge+ hidden representations across the entire model's layers.
        The legend is shown in Fig.~\ref{fig:tsne-xxl}.
    }
    \label{fig:tsne-xxl-full}
\end{figure*}

As shown in Fig.~\ref{fig:tsne-xxl-full}, we provide additional hidden-layer visualizations of USAD~2.0 XXLarge+, complementing Fig.~\ref{fig:tsne-xxl}.
Speech, sound, and music embeddings are already separated into several macro-clusters in the lower layers.
The main difference between lower and upper layers is observed in the speech representations~(shown as squares): lower layers tend to keep phonemes within the same category closer together, whereas upper layers mix different phoneme categories more heavily.
This suggests that the lower layers retain behavior similar to speech SSL models~\cite{chen2022wavlm}, likely due to first-stage SSL distillation, while the upper layers are further aligned with supervised experts through second-stage distillation.

\end{document}